\documentclass[12pt]{article}
\usepackage{latexsym}
\usepackage{amstex}
\usepackage{babel}
\begin{document}
\renewcommand{\thesection}{\Roman{section}.}
\newfont{\fraktgm}{eurm10 scaled 1728}
\newcommand{\graktur}{\baselineskip12.5pt \fraktgm}
\newfont{\fraktfm}{eurm10 scaled 1440}
\newcommand{\frakture}{\baselineskip12.5pt\fraktfm}
\newfont{\fraktrm}{eurm10}
\newcommand{\fraktur}{\baselineskip12.5pt\fraktrm}
\newfont{\fraktem}{eurm6}
\newcommand{\fraktr}{\baselineskip12.5pt\fraktem}
\protect\newtheorem{principle}{Principle}
\newcommand{\dist} {{\rm dist}}
\newcommand{\fini} [1] {{\cal H}_{#1}(\HH)}
\newcommand{\fin} {\fini{fin}} 
\newcommand{\finki} [1] {{\cal K}_{#1}(\HH)}
\newcommand{\fink} {\finki{fin}}
\newcommand{\seta} [1] {I\!\!\!\,#1}
\newcommand{\RR} {{\Bbb R}}
\newcommand{\RRR} {\seta{R}}
\newcommand{\Norm} [1] {\vline \hspace{0.05cm} \vline #1 \vline 
\hspace{0.05cm} \vline \hspace{0.05cm}}
\newcommand{\Betrag} [1] {\vline #1 \vline}
\newcommand{\setb} [1] {I\!\!\!\!\!\:#1}
\newcommand{\setc} [1] {#1\!\!\!#1}
\newcommand{\DD} {\seta{D}}
\newcommand{\NNN} {\seta{N}}
\newcommand{\NN} {{\Bbb N}}
\newcommand{\HH} {{\Bbb H}}
\newcommand{\HHH} {\seta{H}}
\newcommand{\ZZ} {{\Bbb Z}}
\newcommand{\XX} {{\zeta}}
\newcommand{\CC} {{\Bbb C}}
\newcommand{\QQ} {{\Bbb Q}}
\newcommand{\MM} {\seta{M}}
\newcommand{\LL} {\seta{L}}
\newcommand{\FF} {\seta{F}}
\newtheorem{theo}{Theorem}
\newtheorem{nt}{Note}
\newtheorem{lem}{Lemma}
\newtheorem{co}{Corollary}
\newtheorem{de}{Definition}
\newtheorem{fo}{Consequence}
\newtheorem{rle}{Rule}
\newtheorem{rem}{Remark}
\newcommand{\tr} {\mbox{{\rm tr}}}
\newcommand{\norm} [1] {\parallel \hspace{-0.15cm} #1
\hspace{-0.15cm} \parallel}
\newcommand{\betrag} [1] {\mid \hspace{-0.13cm} #1
\hspace{-0.15cm} \mid}
\newcommand{\rsta} [1] {\mid \hspace{-0.1cm} #1 \hspace{-0.04cm}
\rangle}
\newcommand{\lsta} [1] {\langle \hspace{-0.04cm} #1
\hspace{-0.12cm} \mid} 
\newcommand{\pro} {{\cal P}(\HH)} 
\newcommand{\topp} {{\cal T}(\HH)^+_1}
\newcommand{\bop} {{\cal B}(\HH)^+}
\newcommand{\bou} {{\cal B}(\HH)}
\newcommand{\sef} {{\frak E}(\HH)}
\newcommand{\ten} [1] {\otimes_{t \in #1} \HH}
\newcommand{\bout} [1] {{\cal B}^{\otimes}_{#1}(\HH)}
\newcommand{\pout} [1] {{\cal P}^{\otimes}_{#1}(\HH)}
\newcommand{\efut} [1] {{\frak E}^{\otimes}_{#1}(\HH)}
\newcommand{\U} {{\cal U}}
\newcommand{\SSS} {{\cal S}}
\newcommand{\UP} {{\frak U}}
\newcommand{\ieh} {inhomogeneous effect history }
\newcommand{\iek} {inhomogeneous effect history}
\newcommand{\df} {decoherence functional }
\newcommand{\dfs} {decoherence functionals }
\newcommand{\Rea} {\mbox{\em Re }}
\begin{titlepage}
\centerline{\normalsize DESY 96-097  \hfill ISSN 0418 - 9833} 
\centerline{\normalsize May 1996 \hfill} 
\vskip.6in 
\begin{center} 
{\graktur On the Consistent Effect Histories Approach \\ 
\vspace{.15cm}
to Quantum Mechanics} 
\vskip.6in 
{{\Large \frakture Oliver Rudolph} $^*$} 
\vskip.3in 
{\normalsize \sf II.~Institut f\"ur Theoretische Physik, 
Universit\"at Hamburg} 
\vskip.05in 
{\normalsize \sf Luruper Chaussee 149} 
\vskip.05in 
{\normalsize \sf D-22761 Hamburg, Germany}
\vskip.7in
\end{center}
\normalsize
\vfill
\begin{center}
{ABSTRACT}
\end{center}
\smallskip
\noindent
A formulation of the consistent histories approach to quantum 
mechanics in terms of generalized observables (POV measures) and 
effect operators is provided. The usual notion of `history' is 
generalized to the notion of `effect history'.
The space of effect histories carries the structure of a D-poset. 
Recent results of J.D.~Maitland Wright imply that every \df 
defined for ordinary histories can be uniquely extended to a 
bi-additive \df on the space of effect histories. 
Omn\`es' logical interpretation is generalized to the present 
context. The result of this work considerably generalizes and 
simplifies the earlier formulation of the 
consistent effect histories approach to quantum mechanics 
communicated in a previous work of this author. \\ \\
{\small PACS number: 03.65.Bz \\ \\}
\bigskip \noindent
\centerline{\vrule height0.25pt depth0.25pt width4cm \hfill}
\noindent
{\footnotesize $^*$ Internet: rudolph@@x4u2.desy.de} 
\end{titlepage}
\newpage
\section{Introduction}
\indent Nonrelativistic quantum mechanics in its standard 
formulation is not a theory which describes dynamical processes in 
time, but it is a theory which gives probabilities to various 
possible events and measurement outcomes 
at fixed instants of time. The dynamical law of quantum mechanics, 
the Schr\"odinger equation, describes the change of the 
probability amplitude with time. Quantum mechanics in its usual 
form does not provide us with a dynamical law which describes the 
time evolution of events. This can be succinctly summarized by 
saying that quantum mechanics in its usual form does not provide 
us with a (naive) model what is ``actually'' going on on a 
microscopic level in a quantum system. It is often felt that this 
is a serious drawback of quantum mechanics. Examples for attempts 
to modify quantum mechanics to a theory providing us with a model 
for what is ``actually'' happening are hidden variables theories 
(see, e.g., Refs.~\cite{Bell87}, \cite{Bohm87}, 
\cite{Giuntini91}), the dynamical state vector reduction models 
(see, e.g., Refs.~\cite{Ghirardi86,Pearle86,Pearle89} and 
\cite{Gisin89}) or related models (see, e.g., 
Ref.~\cite{Nakano94}). \\
The consistent histories formulation of quantum mechanics is 
another attempt to remedy the situation and to incorporate time 
sequences of events and -- as a special case -- sequential 
measurements into quantum mechanics without providing a naive 
dynamical model for the microscopic world in the above sense and 
without altering the basic principles and the basic mathematical 
structure of Hilbert space quantum mechanics. \\ 
The consistent histories approach to nonrelativistic 
quantum mechanics has been inaugurated in a seminal paper by 
Griffiths \cite{Griffiths84} and further developed by Griffiths 
\cite{Griffiths93,Griffiths94,Griffiths95}, by Omn\`es 
\cite{Omnes88a}-\cite{Omnes95}, by Isham \cite{Isham94}, Isham and 
Linden \cite{IshamL94,IshamL95} and by Isham, Linden and 
Schreckenberg \cite{IshamLS94} and applied to quantum cosmology by 
Gell-Mann and Hartle \cite{GellMann90a}-\cite{GellMann95} and 
Hartle \cite{Hartle91,Hartle94}. Dowker and Kent have carried out 
a critical reexamination of the consistent histories approach and 
particularly of Omn\`es' notion of truth and of the 
Gell-Mann--Hartle programme, see 
Refs.~\cite{Dowker96}-\cite{Kent96}. A critical discussion of the 
consistent histories approach can also be found in 
Ref.~\cite{Zehn.d.}. The consistent histories approach asserts 
that quantum 
mechanics provides a realistic description of individual quantum 
mechanical systems, regardless of whether they are open or closed. 
The possibility of a quantum mechanical description of single 
closed systems, which do neither interact with their environment 
nor are exposed to measurements, is denied by the conventional 
Copenhagen-type interpretations of quantum mechanics. \\
On the contrary, in the logical interpretation developed by 
Omn\`es the notion of measurement is not a key concept. Instead 
one takes the point of view that the aim of an interpretation is 
generally to provide us with a systematic and unambiguous 
language specifying the meaning of the objects in the formalism in 
terms of real physical objects and specifying what can 
meaningfully be said about the physical systems described by the 
theory. 
We will call this attitude the {\em semantic} approach to 
interpretation. Clearly the logical interpretation is a realistic 
interpretation in the sense that it is presupposed that physical 
systems really exist and have real properties regardless of 
whether they are measured or not. \\
A key notion in the formulation of quantum mechanics is the notion 
of observable. In the spirit of the logical interpretation the 
term {\em speakable} would be more appropriate, but we stick to 
the usual terminology. In usual Hilbert space quantum mechanics 
the observables are identified with self-adjoint operators on the 
Hilbert space and propositions about quantum mechanical systems 
are identified with projection operators on Hilbert space. There 
is a one-one correspondence between self-adjoint operators on 
Hilbert space and projection valued (PV) measures on the real line 
$\RR$. To every Borel subset $\cal B$ of $\RR$ there corresponds 
one projection operator representing the proposition that the 
value of the considered observable is in the set $\cal B$. Some 
more remarks about observables and propositions in ordinary 
quantum mechanics can be found in Ref.~\cite{Rudolph96}. \\
The question which objects in the formalism have to be identified 
with observables (or speakables) is clearly a question belonging 
to the interpretation of quantum mechanics. 
Reasonableness and mathematical simplicity are the guiding 
principles to answer this question. The 
most general notion of observable compatible with the 
probabilistic structure of quantum mechanics is that of {\em 
positive operator valued (POV) measures} which contains the 
ordinary observables represented by PV measures on $\RR$ as a 
subclass. Quantum mechanics is totally consistent without POV 
measures, but POV measures enrich the language of quantum 
mechanics and enlarge the measurement theoretical possibilities of 
quantum mechanics \cite{Busch89,Busch91}. On the other hand 
the claim that POV measures 
represent the observables in quantum mechanics is not only 
consistent with the mathematical structure of Hilbert space 
quantum mechanics but furthermore is also reasonable. Many 
examples can be found in the monograph by Busch et al. 
\cite{Busch95}. \\
In this work we take on the view that POV measures are the 
observables in quantum mechanics and that all POV measures should 
be treated on the same footing and that all effects should be 
identified with the general properties (or speakables or beables) 
of quantum systems. We further consider {\em every} effect 
operator as representative of some sort of reality. Some arguments 
supporting this view can be found in Ref.~\cite{Rudolph96} and 
references therein. It is perhaps worthwhile to mention a 
further simple argument which is essentially due to Ludwig 
\cite{Ludwig72}. To this end consider a measuring device $\cal M$ 
consisting of a detector $\cal D$ (designed to measure some 
property $E$ associated with some projection operator) and some 
scatterer $\cal S$. An appropriately prepared incident physical 
system $\cal I$ (e.g., a particle) is first scattered by $\cal S$ 
and then detected by the detector $\cal D$. To obtain the property 
$F$ measured by the device $\cal M$ one has to apply the unitary 
transformation given by the S-matrix $S$ of $\cal S$ to the 
property measured by $\cal D$. Let $\varrho_{\cal I}$ denote the 
initial state of $\cal I$ and $\varrho_{\cal S}$ denote the 
initial state of $\cal S$. Then the relation between $E$ and $F$ 
is given by \[ {\mbox{ tr}} \left(S(\varrho_{\cal I} \otimes 
\varrho_{\cal S})S^{\dagger}(1 \otimes E) \right) = {\mbox{ tr}} 
\left( \varrho_{\cal I} F \right), \]
where the trace on the right hand side is in the Hilbert space 
${\Bbb H}_{\cal I}$ of $\cal I$ and the trace on the left hand 
side is in the tensor product ${\Bbb H}_{\cal I} \otimes 
{\Bbb H}_{\cal S}$ of the Hilbert spaces ${\Bbb H}_{\cal I}$ of 
$\cal I$ and ${\Bbb H}_{\cal S}$ of $\cal S$. The operator $F$ 
is uniquely determined by this equation. However, realistic 
physical S-matrices $S$ transform projection operators (according 
to the above equation) in general to effect operators and only the 
set of effect operators is invariant under this transformations. 
Therefore whether a measuring device 
measures an effect or a property associated with some projection 
operator may depend on an arbitrary cut between the system and the 
apparatus. This argument can be formalized, see 
Ref.~\cite{Kraus83}. \\
In the consistent histories approach it is claimed that all 
results of measurement theory also {\em follow} from the 
consistent histories approach. In the present work we take 
seriously this claim and continue our efforts to formulate the 
consistent histories formalism for general observables represented 
by POV measures. This programme has first been formulated and 
studied in Ref.~\cite{Rudolph96}. We will freely use the 
notation and terminology from Ref.~\cite{Rudolph96} and review 
only the bare essentials. \\
This work is organized as follows: In Section II we summarize the 
consistent histories approach to nonrelativistic Hilbert space 
quantum mechanics and the logical interpretation of quantum 
mechanics. In Section III we recall basic definitions and results 
from Ref.~\cite{Rudolph96} and we formulate our generalized 
(effect) history theory  and a generalized logical rule of 
interpretation for effect histories. Our results are based on an 
important theorem by Wright \cite{Wright95}. This theorem relies 
heavily on the recent solution of the Mackey-Gleason problem, see 
Refs.~\cite{Bunce92,Bunce94}. The results in Section III 
considerably simplify and generalize the results formulated in 
Ref.~\cite{Rudolph96}. 
Section IV presents our summary. \\ As in Ref.~\cite{Rudolph96} it 
must be emphasized that the representation and the 
interpretation of the consistent histories approach in this work 
might not be accepted by the authors cited. 
The present work solely reflects the inclination and the 
views of this author. 
\section{Consistent Histories and the Logical Interpretation}
We consider a quantum mechanical system $\cal S$ without 
superselection rules represented by a separable complex Hilbert 
space $\HH$ and a Hamiltonian operator $H$. Every physical state 
of the considered system is mathematically represented by a 
density operator on $\HH$, i.e., a linear, positive, trace-class 
operator on $\HH$ with trace 1. The time 
evolution is governed by the unitary operator $U(t', t) = 
\exp(-i(t'-t) H / \hbar)$ which maps states at time $t$ into 
states at time $t'$ and satisfies $U(t'', t') U(t',t) = U(t'',t)$ 
and $U(t,t)=1$. \\
In the familiar formulations of quantum mechanics the observables 
are identified with (and represented by) the self-adjoint 
operators on $\HH$ and according to the spectral theorem 
observables can be identified with projection operator valued (PV) 
measures on the real line; 
that is, there is a one-to-one correspondence between self-adjoint 
operators on $\HH$ and maps ${\cal O} : {\cal B}(\RR) \to {\cal 
P}(\HH)$, such that ${\cal O}(\RR) = 1$ and ${\cal O}(\cup_i K_i) 
= \sum_i {\cal O}(K_i)$ for every pairwise disjoint sequence $\{ 
K_i \}_i$ in ${\cal B}(\RR)$ (the series converging in the 
ultraweak topology). Here ${\cal B}(\RR)$ denotes the Borel 
$\sigma$-algebra of $\RR$ and ${\cal P}(\HH)$ denotes the set of 
projection operators on $\HH$, i.e., self-adjoint operators $P$ 
satisfying $P=PP$. \\ 
A meaningful proposition about the system (also called {\em 
physical quality}) is a proposition speci\-fying that the value of 
some observable $\cal O$ lies in some set $B \in {\cal B}(\RR)$. 
This means that to every meaningful proposition about the system 
under consideration there corresponds one projection operator on 
$\HH$. \\ In the state represented by the density 
operator $\varrho$ the probability of a proposition represented by 
the projection operator $P$ is given by ${\tr}(\varrho P)$, where 
$\tr$ denotes the trace in $\HH$. \\ Positive and bounded 
operators $F$ on $\HH$, satisfying $ 0 \leq F \leq 1, $
are commonly called {\sc effects} and the set of 
all effects on the Hilbert space $\HH$ will be denoted by $\sef$. 
We further denote the set of all bounded, linear operators on 
$\HH$ by $\bou$. \\
If $\HH$ is an infinite dimensional Hilbert space, then the set of 
all projection operators $\pro$ on $\HH$ is weakly 
dense in $\sef$ \cite{Davies76}. \\

{\sc Generalized observables} are now identified with {positive 
operator valued (POV) measures} on some measurable space $(\Omega, 
{\cal F}),$ i.e., maps {\fraktur O} $: {\cal F} \to \sef$ with 
the properties: \begin{itemize} \item {\fraktur O}$(A) \geq$ 
{\fraktur O}$(\emptyset)$, for all $A \in \cal F$; \item Let 
$\{A_i\}$ be a countable set of disjoint sets in $\cal F$, then 
{\fraktur O}$(\cup_i A_i) = \sum_i$ {\fraktur O}$(A_i)$, the 
series converging ultraweakly; \item {\fraktur O}$(\Omega) = 1$. 
\end{itemize}
Generalized observables are also called {\sc effect valued 
measures}. 
Ordinary observables (associated with self-adjoint operators on 
$\HH$) are then identified with the projection valued measures on 
the real line $\RR$. Generalizing our above terminology, we regard 
all propositions specifying the value of some generalized 
observable as generalized physical qualities . In 
order to discriminate physical qualities corresponding to ordinary 
observables from physical qualities corresponding to generalized 
observables, we will call the former `ordinary physical 
qualities' and the latter `generalized physical qualities.' In the 
generalized approach to every 
physical quality there corresponds one effect operator. \\
A {\sc homogeneous history} is a map $h : \RR 
\to \pro, t \mapsto h_t$.
We call $t_i(h) := \min(t \in \RR \mid h_t \neq 1)$ the {\sc 
initial} and $t_f(h) := \max(t \in \RR \mid h_t \neq 1)$ the 
{\sc final time } of $h$ respectively. Furthermore, the {\sc 
support of} $h$ is given by ${\frak s}(h) := \{t \in \RR \mid h_t 
\neq 1 \}$. If ${\frak s}(h)$ is 
finite, countable or uncountable, then we say that $h$ is a {\sc 
finite, countable} or {\sc uncountable history} respectively. The 
space of all homogeneous 
histories will be denoted by ${\cal H}(\HH)$, the space of all 
finite homogeneous histories by ${\cal H}_{fin}(\HH)$ and the 
space of all finite homogeneous histories with support $S$ by 
${\cal H}_S(\HH)$. \\
By a {\em history proposition} we mean a proposition about the 
system specifying which history will be realized. We use the terms 
history and history proposition synonymously in this work. \\
In this work we focus attention on finite histories. If a 
homogeneous history vanishes for some $t_0 \in \RR$, i.e., 
$h_{t_0} =0$, then we say that $h$ is a {\sc zero history}. All 
zero histories are collectively denoted by 0, slightly abusing the 
notation.

For every finite subset $S$ of $\RR$ we can consider the Hilbert 
tensor product $\otimes_{t \in S} \HH$ and the algebra $\bout{S} 
:= {\cal B}(\otimes_{t \in S} \HH)$ 
of bounded linear operators on $\ten{S}$. It has been pointed out 
by Isham \cite{Isham94} that for any fixed $S$ there is an 
injective 
(but not surjective) correspondence $\sigma_S$ between finite 
histories with support $S$ and elements of $\bout{S}$ given by 
\begin{equation} \label{E1} \sigma_S : \fini{S} \to \bout{S}, h 
\simeq \{ h_{t_k} \}_{t_k \in S} \mapsto \otimes_{t_k \in S} 
h_{t_k}. \end{equation} 
The finite homogeneous histories with support $S$ can therefore be 
identified with projection operators on $\ten{S}$. The set of 
all projection operators on $\ten{S}$ will in the sequel be 
denoted by $\pout{S}$. However, not all projection operators in 
$\pout{S}$ have the form $\sigma_S(h)$ with $h \in \fini{S}$. \\
The projection operators in $\pout{S}$ are called {\sc finite 
inhomogeneous histories with support $S$} and the space ${\cal 
K}_S(\HH) := \pout{S}$ of projection operators on 
$\ten{S}$ the {\sc space of finite inhomogeneous histories with 
support $S$}. The space of all finite inhomogeneous histories with 
arbitrary support will be denoted by ${\cal K}_{fin}(\HH)$ or by 
$\pout{fin}$. 
Furthermore, to every finite homogeneous history $h \in \fin$ we 
associate its {\sc class 
operator with respect to the fiducial time $t_0$} by $C_{t_0}(h) 
:= U(t_0,t_n) h_{t_n} U(t_n,t_{n-1}) h_{t_{n-1}} ... U(t_2,t_1) 
h_{t_1} U(t_1,t_0)$. 
The class operators can be unambiguously extended to finite 
inhomogeneous histories such that $C_{t_0}$ is additive for 
orthogonal projectors, i.e., 
$C_{t_0}(h \vee k) := C_{t_0}(h) + C_{t_0}(k)$ for $h \perp k$.
The functional $d_{\varrho}:\fink \times \fink 
\to \CC, (h,k) \mapsto d_{\varrho}(h,k) := {\mbox{ tr}} 
\left(C_{t_0}(h) \varrho(t_0) 
C_{t_0}(k)^{\dagger} \right)$ will be called the {\sc 
consistency functional associated with the state} $\varrho$. 
The consistency functional $d_{\varrho}$ satisfies for all $h,h',k 
\in \fink$ \begin{itemize} \item $d_{\varrho}(h,h) \in \RR$ and 
$d_{\varrho}(h,h) 
\geq 0$. \item $d_{\varrho} (h,k) = d_{\varrho}(k,h)^*$. \item 
$d_{\varrho}(1,1) =1$. \item $d_{\varrho}(h \vee h', k) = 
d_{\varrho}(h,k) + d_{\varrho}(h',k),$ whenever $h \perp h'$. 
\item $d_{\varrho}(0,h) =0$, for all 
$h$. \end{itemize} 
In Ref.~\cite{Rudolph96} we have used a slightly different 
terminology: the above consistency functional $d_{\varrho}$ has 
been called there `decoherence functional'. In this work we want 
to carefully distinguish bi-additive functionals on a Boolean 
lattice from bi-additive functionals defined on a D-poset. Thus, 
the former are called consistency functionals, whereas we reserve 
the term `decoherence functional' for a 
bi-additive functional defined on a D-poset (see below). \\ 
Any collection ${\cal C}'$ of histories in $\pout{fin}$ is said to 
be {\sc consistent with respect to the state $\varrho$} if 
${\cal C}'$ is a Boolean algebra (with respect to the 
meet, join and orthocomplementation in $\pout{fin}$ and with unit 
$1_{{\cal C}'}$) and if Re $d_{\varrho}(h,k) = 0$ 
for every two disjoint histories $h, k \in {\cal C}'$.
Here two (possibly inhomogeneous) finite histories $h$ and 
$k$ are said to be {\sc disjoint} if $h \leq \neg k$, where $\leq$ 
is the partial order on ${\cal K}_{{\frak s}(h) \cup {\frak 
s}(k)}(\HH)$. \\ It is now easy to see that the consistency 
functional $d_{\varrho}$ 
induces an additive probability measure $p_{\varrho}$ on every 
consistent Boolean sublattice ${\cal C} \subset \pout{fin}$. 
The probability measure $p_{\varrho}$ is defined by 
\begin{equation} \label{pro} p_{\varrho} : {\cal C} \to \RR^+, 
p_{\varrho}(h) := \frac{d_{\varrho}(h,h)}{d_{\varrho}(1_{\cal C}, 
1_{\cal C})}. \end{equation} 
The probability measure $p_{\varrho}$ on a consistent Boolean 
algebra $\cal C$ of history propositions induced by 
the consistency functional $d_{\varrho}$ according to Equation 
\ref{pro} defines two logical relations in $\cal C$, namely 
an implication and an equivalence relation between histories. A 
history proposition $h$ is said to {\sc imply} a history 
proposition $k$ if the conditional probability $p_{\varrho}(k 
{\mid} h) \equiv \frac{p_{\varrho}(h \wedge_{\cal C} 
k)}{p_{\varrho}(h)}$ is well-defined and equal 
to one. Two history propositions $h$ and $k$ are said to be {\sc 
equivalent} if $h$ implies $k$ and vice versa. \\
The universal rule of interpretation of quantum mechanics can now 
be formulated as 
\begin{rle}[Omn\`es] Propositions about quantum mechanical 
systems should \label{rle1} solely be expressed in terms of 
history propositions. Every description of an isolated quantum 
mechanical system should be expressed in terms of finite history 
propositions belonging to a common consistent Boolean algebra of 
histories. Every reasoning relating several propositions should be 
expressed in terms of the logical relations induced by the 
probability measure from Equation $\mbox{\em \ref{pro}}$ in that 
Boolean algebra. \end{rle}
\section{Consistent Effect Histories}
In Ref.~\cite{Rudolph96} we have motivated and introduced the 
following notion of homogeneous effect history
\begin{de} \label{b2} A {\sc homogeneous effect history (of the 
first kind)} is a map $u : \RR \to {\frak E}(\HH), t \mapsto u_t$.
The {\sc support of} $u$ is given by ${\frak s}(u) := \{t \in \RR 
\mid u_t \neq 1 \}$. If ${\frak s}(u)$ is 
finite, countable or uncountable, then we say that $u$ is a {\sc 
finite, countable} or {\sc uncountable effect history} 
respectively. The space of all homogeneous effect 
histories (of the first kind) will be denoted by ${\Bbb E}(\HH)$, 
the space of all finite homogeneous effect histories (of the first 
kind) by ${\Bbb E}_{fin}(\HH)$ and the space of all finite 
homogeneous effect histories (of the first kind) with support 
$S$ by ${\Bbb E}_S(\HH)$. All homogeneous effect histories for 
which there exists at least one $t \in \RR$ such that $u_t = 0 $ 
are collectively denoted by $0$, slightly abusing the notation. 
\end{de} The class operator 
$C_{t_0}$ defined above for finite ordinary homogeneous histories 
can be defined for homogeneous finite effect histories $u \in 
{\Bbb E}_{fin}(\HH)$ \begin{eqnarray*} \label{cop}
C_{t_0}(u) & := & U(t_0,t_n) \sqrt{u_{t_n}} 
U(t_n,t_{n-1}) \sqrt{u_{t_{n-1}}} ... U(t_2,t_1) 
\sqrt{u_{t_1}} U(t_1,t_0). \end{eqnarray*} For every pair 
$u$ and $v$ of finite homogeneous effect histories (of the first 
kind) we define the {\sc decoherence weight} of $u$ and $v$ by \[ 
d_{\varrho} (u,v) := {\tr} \left(C_{t_0}(u) \varrho(t_0) 
C_{t_0}(v)^{\dagger} \right). \] The functional $d_{\varrho}: 
{\Bbb E}_{fin}(\HH) \times {\Bbb E}_{fin}(\HH) 
\to \CC, (u,v) \mapsto d_{\varrho}(u,v)$ will be called the {\sc 
decoherence functional associated with the state} $\varrho$. \\
The map $\sigma_S$ given by Equation \ref{E1} can be extended to a 
map \begin{equation} \label{E2} \sigma_{fin} : {\Bbb E}_{fin}(\HH) 
\to \bout{fin}, u \simeq \{ u_{t_k} \}_{t_k \in {\frak s}(u)} 
\mapsto \otimes_{t_k \in {\frak s}(u)} u_{t_k}, \end{equation} 
where $\bout{fin}$ denotes the disjoint union of all $\bout{S}, S 
\subset \RR$ finite. 
The map $\sigma_{fin}$ is neither injective nor surjective. 
However, $d_{\varrho}(u,v)$ depends on $u$ and $v$ only through 
$\sigma_{fin}(u)$ and $\sigma_{fin}(v)$. 
From a mathematical point of view it thus seems to be natural to 
define the notion of {\em inhomogeneous} effect history as 
follows:
\begin{de} Let $S$ be a finite subset of $\RR$, then we call the 
space $\efut{S} := {\frak E}(\ten{S})$ of effect operators on 
$\ten{S}$ the {\sc space of finite inhomogeneous effect histories 
with support $S$}. The space of all finite inhomogeneous effect 
histories with arbitrary support will be denoted by $\efut{fin}$. 
The elements in $\efut{fin}$ will also be called {\sc effect 
history propositions}. \label{D2} \end{de} 
The homogeneous elements in $\efut{fin}$ represent equivalence 
classes of homogeneous effect histories. In this work we will 
carefully distinguish between homogeneous effect histories as 
defined in Definition \ref{b2} and homogeneous elements in 
$\efut{fin}$. For clarity of exposition we will call the former 
{\em homogeneous effect history of the first kind} or (where no 
confusion can arise) simply homogeneous effect histories, whereas 
the latter will be called {\em homogeneous effect histories of the 
second kind}. \\ 
In technical terms $\efut{fin}$ is the direct limit of the 
directed system $\{\efut{S} \mid S \subset \RR \mbox{ finite} \}$.
All the $\efut{S}$, $S \subset \RR$ and $\efut{fin}$ carry several
distinct D-poset structures, as discussed in 
Ref.~\cite{Rudolph96}. We will denote the partial addition on 
$\efut{fin}$ by $\oplus$ and the partial substraction on 
$\efut{fin}$ by $\ominus$. For further literature on D-posets and 
effect algebras see Refs.~\cite{Foulis94}-\cite{Pulmannova95}. As 
in Ref.~\cite{Rudolph96} we use the terms {\em D-poset} and {\em 
effect algebra} synonymously. We refer to the D-poset structure on 
$\efut{fin}$ given by the partial addition $\oplus$, where $E_1 
\oplus E_2$ is defined if $E_1 + E_2 < 1$ by $E_1 \oplus E_2 := 
E_1 + E_2$, as the {\sc canonical D-poset structure}. \\ 
In Ref.~\cite{Rudolph96} we have used a different notion of \iek, 
because it was not 
clear whether the \df $d_{\varrho}$ defined above on the space 
of homogeneous effect histories (of the first kind) can be 
(uniquely) extended to an in some appropriate sense bi-additive 
functional on the space of inhomogeneous effect histories as 
defined in Definition \ref{D2}. In Ref.~\cite{Rudolph96} we gave a 
rather technical definition the notion of inhomogeneous effect 
history. Essentially we defined an inhomogeneous effect history to 
be a member of the free lattice generated by the homogeneous 
effect histories (of the first kind) by at most finitely many 
applications of the grammatical connectives 'and' and 'or', to 
wit, we have viewed inhomogeneous effect histories to be -- in 
essence -- propositions in the language of quantum mechanics 
involving several (but at most finitely many) homogeneous effect 
histories. In turn only the latter were viewed as the basic {\em 
physical} entities in the formalism. We have shown in 
Ref.~\cite{Rudolph96} that with this definitions it is possible to 
consistently extend the consistent histories formulation of 
quantum mechanics and to incorporate effect histories. However, 
this approach involves a rather technical and mathematically by no 
means canonical definition of the notion of inhomogeneous history. 
Inhomogeneous effect histories in the sense of 
Ref.~\cite{Rudolph96} represent only semantical entities without 
an obvious physical interpretation. In this work the term \ieh is 
always meant in the 
sense of Definition 2 unless explicitly otherwise stated.  \\
In this work we use a recent result of J.D.~Maitland Wright 
\cite{Wright95} 
which implies that the \df $d_{\varrho}$ as defined above on the 
space of homogeneous effect histories (of the second kind) can 
indeed be extended to a functional on the space of inhomogeneous 
effect histories with the desired properties. We first recall the 
central result from Ref.~\cite{Wright95} 
\begin{theo} Let $A$ be a von Neumann algebra with no type 
$I_2$ direct summand. Let $d : {\cal P}(A) \times {\cal P}(A) \to 
\CC$ be a \df. If $d$ is bounded, then $d$ extends to a unique 
bounded bilinear functional $\widetilde{d}$ on $A \times A$. 
Furthermore, $d$ is continuous when ${\cal P}(A)$ is equipped with 
the topology induced by the norm of $A$. Also, $d(u,v)^* = d(v^*, 
u^*)$. \end{theo} 

If $A$ is a von Neumann algebra, let ${\cal P}(A)$ denote the set 
of projectors in $A$. A function $d : {\cal P}(A) 
\times {\cal P}(A) \to \CC$ is called a {\sc \df}, if (i) $d(p_1 
\oplus p_2, q) = d(p_1,q) + d(p_2,q),$ whenever $p_1$ and $p_2$ 
are mutually orthogonal; (ii) $d(p,q)^* = d(q,p)$; (iii) $d(p,p) 
\geq 0$; (iv) $d(1,1) = 1$. 

Since the set ${\cal B}({\HH})$ of bounded operators on a Hilbert 
space $\HH$ with dimension greater than 2 is a von Neumann algebra 
(of type $I$), Theorem 1 can be applied to the \df $d_{\varrho} : 
\fink \times \fink \to \CC, (h,k) \mapsto d_{\varrho}(h,k) := $ 
tr $\left(C_{t_0}(h) \varrho(t_0) C_{t_0}(k)^{\dagger} 
\right)$ defined in Section II above. Thus for every finite subset 
$S \subset \RR$ there is a unique bounded bilinear functional 
$\widetilde{d}_{\varrho,S}$ on $\bout{S} \times \bout{S}$ 
extending the \df $d_{\varrho}$ restricted to $\pout{S} \times 
\pout{S}$. \\

The restriction $\widehat{d}_{\varrho,S}$ of 
$\widetilde{d}_{\varrho,S}$ 
to $\efut{S} \times \efut{S}$ is a bounded functional which is 
additive in both arguments with respect to the canonical D-poset 
structure on $\efut{S}$. The collection of all such functionals 
$\widehat{d}_{\varrho,S}$ for any finite $S \subset \RR$ induces 
a bounded functional $\widehat{d}_{\varrho}$ on $\efut{fin} \times 
\efut{fin}$ which is additive in both arguments with respect to 
the canonical D-poset structure on $\efut{fin}$. The functional 
$\widehat{d}_{\varrho}$ will be called the {\sc \df with respect 
to the state $\varrho$ on $\efut{fin}$}. 

Since $\efut{fin}$ is a D-poset, $\efut{fin}$ is in particular a 
partially ordered set. However, for two elements 
$e_1, e_2 \in \efut{fin}$ the supremum $e_1 \vee e_2$ and the 
infimum $e_1 \wedge e_2$ not necessarily exist, that is, 
$\efut{fin}$ is not a lattice. But there exists a partially 
defined join operation denoted by $\vee$ and a partially defined 
meet operation denoted by $\wedge$. To every element $e \in 
\efut{fin}$ there exists one unique element $e' \in \efut{fin}$ 
such that $e \oplus e'$ is well-defined and $e \oplus e' = 1$. We 
refer to $e' = 1 \ominus e$ as to the {\sc complement of } $e$. 
\begin{de} \label{CEH} A subset ${\cal B} \subset \efut{fin}$ 
is said to be an {\sc admissible Boolean lattice of 
(inhomogeneous) effect histories} 
if the following conditions are satisfied \begin{itemize} \item 
There exist two binary operations on $\cal B$, denoted by 
$\vee_{\cal B}$ and $\wedge_{\cal B}$ respectively, and one unary 
operation on $\cal B$, denoted by $\neg_{\cal B}$, such that 
the operations $\vee_{\cal B}, \wedge_{\cal B}$ and $\neg_{\cal 
B}$ are compatible with the partial order on $\cal B$ induced by 
the partial order on $\efut{fin}$ and such that $({\cal B}, 
\vee_{\cal B}, \wedge_{\cal B}, \neg_{\cal B})$ is a 
Boolean lattice. I.e., $\vee_{\cal B}$ is the join operation, 
$\wedge_{\cal B}$ is the meet operation and $\neg_{\cal B}$ is the 
complementation operation on $\cal B$; the lattice-operations 
$\vee_{\cal B}$ and $\wedge_{\cal B}$ coincide with the 
partially defined meet operation $\vee$ and join operation 
$\wedge$ on $\efut{fin}$ whenever the latter are well-defined, to 
wit, $e_1 \wedge_{\cal B} e_2 = e_1 \wedge e_2$ and $e_3 
\vee_{\cal B} e_4 = e_3 \vee e_4$ for all $e_1,e_2,e_3,e_4 \in 
\cal B$, whenever the right hand sides are well-defined in 
$\efut{fin}$. 
The lattice-operations $\vee_{\cal B}$ and $\wedge_{\cal B}$ are 
such that a complementation $\neg_{\cal B}$ can be unambiguously 
defined on $\cal B$; 
\item There exists an injective map ${\frak M} : 
{\cal B} \to \efut{fin}$ 
which satisfies the following conditions \begin{itemize} \item 
$\frak M$ is a {\sf positive valuation} on $\cal B$ 
with values in $\efut{fin}$, to wit, a map satisfying the {\tt 
valuation condition } ${\frak M}\left({b}_1 \vee_{{\cal B}} {b}_2 
\right) \ominus {\frak M}({b}_1)= {\frak M}({b}_2) \ominus {\frak 
M}({b}_1 \wedge_{{\cal B}} {b}_2)$, for all 
${b}_1, {b}_2 \in {\cal B}$. This condition means in particular 
that the left hand side and the right hand side are well-defined 
for all ${b}_1, {b}_2 \in {\cal B}$; \item $\frak M$ preserves 
decoherence weights, i.e., $\widehat{d}_{\varrho}(e_1, e_2) = 
\widehat{d}_{\varrho}({\frak M}(e_1), {\frak M}(e_2))$, for all \\ 
$e_1, e_2 \in \cal B$. \end{itemize} \end{itemize} 
An admissible Boolean sublattice of $\efut{fin}$ will be briefly 
denoted by $({\cal B}, {\frak M})$. \end{de} 
\begin{rem} Strictly speaking a {\sc sublattice $\cal L$ 
of} $\efut{fin}$ is a subset ${\cal L} \subset \efut{fin}$ 
such that $\cal L$ endowed with the restrictions of $\vee$ and 
$\wedge$ to $\cal L$ is a lattice. It makes thus sense to speak of 
sublattices of $\efut{fin}$. However, it is important to notice 
that an admissible Boolean sublattice of $\efut{fin}$ is not 
necessarily a sublattice of $\efut{fin}$ in this sense. \end{rem}
\begin{rem} Let ${\cal L}_1, {\cal L}_2$ be lattices. A map $\nu : 
{\cal L}_1 \to {\cal L}_2$ is called {\sc positive} if $\nu(p) < 
\nu(q)$, whenever $p < q$. Since any Boolean lattice is relatively 
complemented, the condition that the map $\frak M$ in Definition 
\ref{CEH} is positive is actually redundant and follows already 
from the Definition of D-posets and from the valuation condition. 
In particular $\frak M$ is order preserving. \end{rem} 
\begin{rem} The complement $e' = 1-e$ in $\efut{fin}$ of some 
element $e \in \cal B$ does in general not coincide with the 
complement $\neg_{{\cal B}} e$ in $\cal B$. 
The greatest element $1_{\cal B}$ and the least 
element $0_{\cal B}$ in $\cal B$ do not necessarily coincide with 
the greatest element $1$ and the least element 
$0$ in $\efut{fin}$ respectively. \end{rem} 
Our target is to generalize Omn\`es' logical rule and thus to 
single out the appropriate subsets of $\efut{fin}$ on which a 
reasoning involving (inhomogeneous) effect histories compatible 
with `common sense' can be defined. The conditions in Definition 
\ref{CEH} are clearly the minimal structure required. Usually 
`common sense' (compare Ref.~\cite{Omnes94}) is tacitly associated 
with Boolean lattices. 
Thus the first condition in Definition \ref{CEH} that $\cal B$ is 
a Boolean lattice is indispensable. We have already mentioned 
above that the set $\efut{fin}$ carries (amongst others) a 
canonical D-poset structure, but no lattice structure and that 
the \df $\widehat{d}_{\varrho}$ is additive with respect to the 
D-poset structure on $\efut{fin}$. In the consistent histories 
approach, however, reasoning is defined on Boolean lattices $\cal 
B$ with the help of consistency functionals which are additive 
with respect to the 
lattice structure of $\cal B$. Thus one has to restrict oneself to 
Boolean lattices ${\cal B} \subset \efut{fin}$ such that the 
lattice structure of $\cal B$ is exactly mirrored in the D-poset 
structure of $\efut{fin}$ (by the map $\frak M$). This leads to 
the condition that there exists a positive valuation $\frak M$ as 
required in the second condition of Definition \ref{CEH}. 
The reasoning to be defined should be independent of the map 
$\frak M$ chosen. Thus it is necessary to require that $\frak M$ 
preserves decoherence weights. 
\begin{rem} \label{rem11} The decoherence functional 
$\widehat{d}_{\varrho}$ induces a consistency functional 
$d_{\varrho, {\cal B}}$ on ${\cal B} 
\times {\cal B}$ by $d_{\varrho, {\cal B}} : {\cal B} \times 
{\cal B} \to \CC, d_{\varrho, {\cal B}}(p_1, p_2) := 
\widehat{d}_{\varrho} ({\frak M}(p_1),{\frak M}(p_2))$, which is 
additive in both arguments with respect to the Boolean lattice 
structure on $\cal B$. \end{rem} 
\begin{de} An admissible Boolean lattice $({\cal B}, {\frak M})$ 
is called {\sc consistent w.r.t.~$\varrho$} if for every pair of 
disjoint elements $b_1, b_2 \in \cal B$ (i.e., elements satisfying 
$b_1 \wedge_{\cal B} b_2 = 0$) the {\tt consistency 
condition } $\mbox{{\em Re }} d_{\varrho, {\cal B}}(b_1, b_2) = 
0$ is satisfied. \end{de} 

\begin{theo} \label{th1} Let $({\cal B}, {\frak M})$ be a 
consistent admissible Boolean lattice of effect histories. Then 
the consistency functional $d_{\varrho, {\cal B}}$ induces a 
probability functional $p_{\varrho, {\cal B}}$ on ${\cal B}$ by $b 
\mapsto p_{\varrho, {\cal B}}(b) \equiv \frac{d_{\varrho, {\cal 
B}}({\frak M}(b), {\frak M}(b))}{d_{\varrho, {\cal B}}({\frak 
M}(1_{\cal B}), {\frak M}(1_{\cal B}))}$. \end{theo}

\begin{de} An effect history proposition $e_1 \in \efut{fin}$ is 
said to {\sc imply} an effect history proposition $e_2 \in 
\efut{fin}$ in the state $\varrho$ if there exists a consistent 
admissible Boolean sublattice $\cal B$ of $\efut{fin}$ containing 
$e_1$ and $e_2$ \label{D6} and if the conditional probability 
$p_{\varrho, {\cal B}}(e_2 {\mid} e_1) \equiv \frac{p_{\varrho, 
{\cal B}}(e_1 \wedge_{\cal B} e_2)}{p_{\varrho, {\cal B}}(e_1)}$ 
is well-defined and equal to one. We write $e_1 
\Longrightarrow_{\varrho} e_2$. 
Two history propositions $e_1$ and $e_2$ are said to be {\sc 
equivalent} if $e_1$ implies $e_2$ and vice versa. We write $e_1 
\Longleftrightarrow_{\varrho} e_2$. \end{de}
\begin{rem} If $e_1 \wedge e_2$ exists in $\efut{fin}$, then it is 
easy to verify that if $p_{\varrho, {\cal B}_0}(e_2 \mid e_1)$ is 
well-defined and equal to one in some consistent admissible 
Boolean lattice ${\cal B}_0$ containing $e_1$ and $e_2$, then 
$p_{\varrho, {\cal B}}(e_2 \mid e_1)$ is well-defined and equal to 
one in every consistent admissible Boolean lattice $\cal B$ 
containing $e_1$ and $e_2$. If $e_1 \wedge e_2$ does not exists in 
$\efut{fin}$, then there may be consistent Boolean lattices ${\cal 
B}_1$ containing $e_1$ and $e_2$ such that $p_{\varrho, {\cal 
B}_1}(e_2 \mid e_1)$ is not one or is not well-defined. If $e_1 
\wedge e_2$ does not exists in $\efut{fin}$, then it seems 
reasonable to define $e_1 \Longrightarrow_{\varrho} e_2$ if there 
exists an admissible Boolean lattice $\cal B$ containing $e_1, 
e_2$ and some further element $e_3 \in \efut{fin}$ satisfying $e_1 
\geq e_3$ and $e_2 \geq e_3$ such that $\frac{p_{\varrho, {\cal 
B}}(e_3, e_3)}{p_{\varrho, {\cal B}}(e_1, e_1)}$ is well-defined 
in $\cal B$ and equal to one. \end{rem} 

The generalized universal rule of interpretation of quantum 
mechanics can now simply be formulated as 
\addtocounter{rle}{1}
\begin{rle} Propositions about quantum mechanical 
systems should \label{rle2} solely be expressed in terms of 
effect history propositions. Every description of an isolated 
quantum mechanical system should be expressed in terms of finite 
effect history propositions belonging to a common consistent 
admissible Boolean algebra of effect histories. Every reasoning 
relating several propositions should be 
expressed in terms of the logical relations induced by the 
probability measure from Theorem $\mbox{\em \ref{th1}}$ in that 
Boolean algebra. \end{rle} 

(This rule is numbered 'Rule \ref{rle2}' in order to distinguish 
it from Rule 2 stated in Ref.~\cite{Rudolph96}.) It is instructive 
to compare Rule \ref{rle2} with Rule 2 stated in 
Ref.~\cite{Rudolph96}. 
It is obvious that Rule \ref{rle1} is contained in Rule \ref{rle2} 
as a special case. A more extensive discussion of the motivation 
and the philosophy underlying the logical interpretation of 
quantum mechanics can be found in 
Refs.~\cite{Omnes90}-\cite{Omnes94} and in Ref.~\cite{Rudolph96} 
and will not be repeated here. \\
Compared with the treatment in Ref.~\cite{Rudolph96} we have 
achieved a 
considerable simplification of the logical interpretation in terms 
of generalized observables and of the formalism of the consistent 
effect histories approach to generalized quantum mechanics.
From a mathematical point of view, the extension of the ordinary 
consistent histories approach given in this article is a natural 
one. \\
Rule \ref{rle2} asserts that to every meaningful proposition about 
a quantum mechanical system there is an inhomogeneous effect 
history $e \in \efut{fin}$. However, homogeneous effect histories 
of the first kind, which have a direct physical interpretation,  
are not contained in $\efut{fin}$. According to Rule \ref{rle2} 
homogeneous effect histories of 
the first kind can only indirectly be included into a description 
of a quantum mechanical system by representing every homogeneous 
effect history of the first kind $e$ by its corresponding 
homogeneous effect history of the second kind $\sigma_{fin}(e)$. 
\\ It remains to determine the connection of Rule \ref{rle2} 
stated above and the generalized logical rule (Rule 2) formulated 
in Ref.~\cite{Rudolph96}. In contrast to Rule \ref{rle2} above, 
the propositions about a quantum mechanical system permitted by 
Rule 2 stated in Ref.~\cite{Rudolph96} contain the homogeneous 
effect histories 
of the first kind as a subclass and accordingly a description of a 
quantum mechanical system and reasoning can be done directly in 
terms of homogeneous effect histories of the first kind. In the 
next subsection we will see, however, that in an appropriate sense
Rule \ref{rle2} is a generalization of Rule 2 stated in 
Ref.~\cite{Rudolph96} and that a description and reasoning 
(permitted by Rule 2) 
directly in terms of homogeneous effect histories of the first 
kind can always be lifted to a description and reasoning 
(permitted by Rule \ref{rle2}) in terms of the corresponding 
homogeneous effect histories of the second kind.  
\subsection*{The connection between admissible and allowed Boolean 
lattices}
In this subsection we will show that Rule \ref{rle2} formulated 
above is indeed a generalization of the generalized logical rule 
as formulated in Ref.~\cite{Rudolph96}. In this subsection we will 
use the notation and terminology introduced in 
Ref.~\cite{Rudolph96} without further notice. In this subsection 
the term {\em homogeneous effect history} is always meant to 
denote homogeneous effect histories of the first kind.

Consider some homogeneous effect history of order $k > 0$ denoted 
by $w_{E_1,..., E_m}^k$, where $E_1, ..., E_m \in \sef$. The 
corresponding history proposition states that first at $k$ 
successive times $t_{1,1}, ..., t_{1,k}$ the appropriately time 
translated effect $E_1(t_{1,j}) = U(t_{1,j}, t_{1,1}) E_1 
U(t_{1,j}, t_{1,1})^{\dagger}$ ($1 \leq j \leq k$) is realized and 
then at $k$ successive times $t_{2,1}, ..., t_{2,k}$ the effect 
$E_2(t_{2,j}) = U(t_{2,j}, t_{1,1}) E_2 U(t_{2,j}, 
t_{1,1})^{\dagger}$ ($1 \leq j \leq k$) and so on (we refer the 
reader to the discussion following Theorem 4 in 
Ref.~\cite{Rudolph96}; 
for simplicity we assume that the history $w_0$ appearing there is 
the unit history, i.e., $(w_0)_t = 1$ for all $t$). \\
Now we first observe that the exact times associated with the 
effects in some homogeneous effect history are inessential. The 
only thing that physically matters is the order and sequence of 
the effects in the homogeneous history. The time points associated 
with the effect operators in some homogeneous effect history can 
be changed provided the order remains fixed and provided the 
effect operators associated with the shifted times are 
appropriately time translated with the unitary evolution operator 
$U$. We say that two homogeneous effect histories related in this 
way to each other are {\sc shift-equivalent}. \\ 
If we define $F_j := E_j^{k/2}$, for all $1 \leq j \leq k$, then 
we see that every homogeneous effect history $w^k_{E_1, ..., E_m}$ 
of order $k$ can be mapped to a homogeneous effect history 
$w^2_{F_1, ..., F_m}$ of order 2. This map preserves decoherence 
weights. The history $w^2_{F_1, ..., F_m}$ is unique up to 
shift-equivalence. That the $F_j$ are effect operators follows 
from Proposition 2 in Ref.~\cite{Langer62}. We further recall that 
$F \oplus_1 F' = F \oplus F' = \left( E \oplus_{2/k} E' 
\right)^{k/2}$ where $F = E^{k/2}$ and $F' = (E')^{k/2}$ whenever 
the expressions are well-defined. Now it is easy to see that for 
every allowed Boolean algebra ($\cal B$, {\fraktur B}) of order 
$k$ (as defined in Ref.~\cite{Rudolph96}) there exists an allowed 
Boolean algebra ($\cal B$', {\fraktur B}') of order 2 and a 
lattice isomorphism $\varphi: {\cal B} \to {\cal B}'$ such that 
{\fraktur B} = {\fraktur B}'$ \circ \varphi$. Thus, it suffices 
to consider allowed Boolean algebras of order 2 in the sequel. 
In Theorem \ref{T3} below $\frak N$ denotes the canonical map 
defined in Remark 15 in Ref.~\cite{Rudolph96}. 
\begin{theo} Let $({\cal B}, {\frak M})$ be an admissible Boolean 
lattice in the sense of Definition \ref{CEH} above and let $({\cal 
A}, {\frak J})$ be an allowed Boolean lattice of effect histories 
of order $k$ as defined in Ref.~\cite{Rudolph96}. Let ${\cal A}_0$ 
denote the set of atoms of $\cal A$. Then there exists a 
lattice isomorphism $\psi : {\cal A} \to \cal B$ preserving 
decoherence weights and satisfying ${\frak J} = {\frak M} \circ 
\psi$ if and only if $\cal B$ is atomic, and $\frak M$ maps the 
set ${\cal B}_0$ of atoms of $\cal B$ bijectively to ${\frak 
J}({\cal A}_0)$, and ${\frak M}(0_{\cal B}) = {\frak J}(0_{\cal 
A})$. \label{T3} \end{theo} 
{\bf Proof:} $"\Longrightarrow"$: trivial. $"\Longleftarrow"$: 
${\frak M}^{-1} \circ {\frak J}$ restricted to $\widetilde{\cal A} 
:= {\cal A}_0 \cup \{ 0_{\cal A} \}$ can in 
an obvious way be extended to a lattice isomorphism $\psi : {\cal 
A} \to {\cal B}$ by requiring $\psi( \vee_{{\cal A}, i \in I} a_i) 
= \vee_{{\cal B}, i \in I} \psi(a_i)$ for any $\{a_i\}_{i \in I} 
\subset {\cal A}_0$. Then $\widetilde{{\frak J}} := {\frak M} 
\circ \psi$ is a positive valuation satisfying the valuation 
condition and extending the map ${\frak N}_{\mid \widetilde{\cal 
A}}$ as required in 
the Definition of the allowed Boolean lattice. Since $({\cal A}, 
{\frak J})$ is an allowed Boolean lattice, $\frak J$ is the unique 
positive valuation with this property and thus ${\frak J} = 
\widetilde{\frak J}$. That $\psi$ preserves decoherence weights 
follows immediately: $d_{\varrho, {\cal A}}(a_1, a_2) := 
d_{\varrho, {\frak T}, k}({\frak J}(a_1), {\frak J}(a_2)) = 
\widehat{d}_{\varrho}({\frak M} \circ \psi (a_1), {\frak M} \circ 
\psi(a_2)) = \widehat{d}_{\varrho}(\psi(a_1), \psi(a_2))$, for all 
$a_1, a_2 \in \widetilde{\cal A}$, where $d_{\varrho, {\frak T}, 
k}$ denoted the \df on ${\frak E}(\HH)_{2/k, {\frak T}}$ (compare 
Remark 20 in Ref.~\cite{Rudolph96}). \hfill $\Box$ \\ 
\begin{theo} For every allowed Boolean lattice $({\cal A}, {\frak 
J})$ in the sense of Ref.~\cite{Rudolph96} there is an admissible 
Boolean lattice $({\cal B}, {\frak M})$ such that there exists an 
isomorphism $\psi : {\cal A} \to {\cal B}$ satisfying the 
conditions from Theorem \ref{T3}. \label{T4} \end{theo}
{\bf Proof}: We denote by ${\cal A}_0$ the set of atoms of $\cal 
A$. We construct $\cal B$ inductively. We choose ${\frak J}({\cal 
A}_0)$ to be the set of atoms of 
$\cal B$ and $0_{\cal B} := {\frak J}(0_{\cal A})$. We define 
${\frak J}(a_1) \vee_{\cal B} {\frak J}(a_2) 
:= {\frak J}(a_1 \vee_{\cal A} a_2)$ for all $a_1, a_2 \in {\cal 
A}_0$. If ${\cal A}_0$ contains more than two elements, then 
${\frak J}(a_1) \vee_{\cal B} {\frak J}(a_2) = {\frak 
J}(a_1) \oplus {\frak J}(a_2)$  for $a_1 \neq a_2$ . This 
definition makes sense since ${\frak J}(a_1) \oplus 
{\frak J}(a_2)$ is well-defined for all $a_1, a_2 \in {\cal A}_0$ 
with $a_1 \neq a_2$ and since ${\frak J}(a_1) \neq {\frak J}(a_2)$ 
for all $a_1, a_2 \in {\cal A}_0$ with $a_1 \neq a_2$. If ${\cal 
A}_0$ contains exactly two elements, then ${\frak J}(a_1) 
\vee_{\cal B} {\frak J}(a_2) = {\frak J}(a_1) \oplus {\frak 
J}(a_2) \ominus {\frak J}(0_{\cal A})$ for $a_1 \neq a_2$. This 
definition makes sense since ${\frak J}(a_1) \oplus 
{\frak J}(a_2) \ominus {\frak J}(0_{\cal A})$ is well-defined for 
all $a_1, a_2 \in {\cal A}_0$ with $a_1 \neq a_2$ and since 
${\frak J}(a_1) \neq {\frak J}(a_2)$ for all $a_1, a_2 \in {\cal 
A}_0$ with $a_1 \neq a_2$. \hfill $\Box$ \\ \\
The {\em full D-posets} also discussed in Ref.~\cite{Rudolph96} 
are trivially contained in the class of admissible Boolean 
lattices defined in Definition \ref{CEH}. \\
From Theorem \ref{T4} and our Definition \ref{D6} of the 
implication relation between effect histories it follows 
immediately that if $e_1$ and $e_2$ are homogeneous effect 
histories such that $e_1 \Longrightarrow_{\varrho} e_2$ in the 
sense of Ref.~\cite{Rudolph96}, then also $e_1 
\Longrightarrow_{\varrho} 
e_2$ in the sense of Definition \ref{D6}. \\
Thus, Theorem \ref{T4} clearly shows that Rule \ref{rle2} is 
indeed a generalization of the Rule 2 stated in 
Ref.~\cite{Rudolph96}. 

\section{Summary}
We now summarize our discussion by stating the general axioms for 
a generalized quantum theory based on our generalized history 
concept. This subsection parallels the discussion in 
Ref.~\cite{Isham94}. \begin{enumerate}
\item The space $\frak U$ of general 
history propositions.
\begin{itemize} \item The space $\frak U$ carries a canonical 
D-poset structure denoted by $\oplus$. \begin{itemize} \item In 
this work $\frak U$ is given by $\efut{fin}$. \end{itemize} 
\end{itemize} 
\item The space $\cal U$ of history filters or homogeneous 
histories. \begin{itemize} \item $\cal U$ is the space of the 
basic physical properties of a physical system with a direct 
physical interpretation. An element of $\cal U$ is a time-ordered 
sequence of one-time propositions about the system. There exists a 
map $F$ mapping the elements of $\cal U$ to a D-poset $\frak E$. 
$\frak E$ can be interpreted as the set of (equivalence classes 
of) one-time propositions. \begin{itemize} \item In this work 
$\cal U$ equals the space of homogeneous effect histories of the 
first kind ${\cal U} = {\Bbb E}_{fin}(\HH)$, cf.~Definition 
\ref{b2}; ${\frak E}$ is 
given by ${\frak E}(\HH)$ and $F$ is given by $F(u) = 
C_{t_0}(u)^{\dagger}C_{t_0}(u)$. \end{itemize} \item $\cal U$ 
is a partially ordered set with unit history 1 and null history 0. 
\item There exists an order preserving map $\tau : {\cal U} \to 
\frak U$, i.e., $\tau({\cal U}) \subset \frak U$. \begin{itemize} 
\item In this work $\tau$ is given by $\sigma_{fin}$. 
\end{itemize} \item $\cal U$ is a partial semigroup with 
composition law $\circ$, cf.~Ref.~\cite{Isham94}. $a \circ b$ is 
well-defined if $t_f(a) 
< t_i(b)$. In this case we say that $a$ {\sc proceeds} $b$ or that 
$b$ {\sc follows} $a$. Further, $1 \circ a = a \circ 1 = a$ and $a 
\circ 0 = 0 \circ a =0$. If $a \circ b$ is defined, then $a\circ 
b = a \wedge b$, in particular the right hand side is 
well-defined. \item The partial ordering on $\cal U$ induces a 
partial unary operation $\neg$ (complementation) and two 
partial binary operations $\wedge$ and $\vee$ (meet and join) on 
${\cal U}$. \end{itemize} \pagebreak[3] 
\item The space of decoherence functionals. 
\begin{itemize} \item 
A decoherence functional is a map $d : {\frak U} \times 
{\frak U} \to \CC$ which satisfies for all $\alpha, 
\alpha',\beta \in {\frak U}$
\begin{itemize} \item $d(\alpha,\alpha) \in \RR$ and 
$d(\alpha,\alpha) 
\geq 0$. \item $d(\alpha,\beta) = d(\beta,\alpha)^*$. \item 
$d(1,1) =1$. \item $d(0,\alpha) =0$, for all 
$\alpha$. \item $d(\alpha_1 \oplus 
\alpha_2, \beta) = d(\alpha_1,\beta) + d(\alpha_2,\beta)$ for all 
$\alpha_1, \alpha_2, \beta \in {\frak U}$ for which 
$\alpha_1 \oplus \alpha_2$ is well-defined. \end{itemize}
\item In Ref.~\cite{Rudolph96} it was possible to explicitly 
construct the \df on all inhomogeneous effect histories 
considered. In this work we have no explicit construction of the 
\df on $\efut{fin}$. Only its existence is known by Theorem 1. 
\end{itemize} 
\item The physical interpretation.
\begin{itemize} \item The physically interesting subsets of $\frak 
U$ are the `admissible' Boolean sublattices ${\cal B}$ of ${\frak 
U}$ (see Definition \ref{CEH}) on which a positive valuation 
$\frak M$ can be defined with values in $\frak U$ such that for 
every $u \in {\cal B}$ the value ${\frak M}(u)$ does not depend 
upon the particular `admissible' Boolean lattice ${\cal B}$ 
chosen. 
\item The map $\frak M$ `lifts' the lattice structure of $\cal B$ 
to the D-poset structure of $\frak U$ and every \df on $\frak U$ 
induces a consistency functional on $\cal B$.  
\item The \df induces a probability measure on the 
consistent (w.r.t.~the decoherence functional) `admissible' 
Boolean sublattices of ${\frak U}$. \item On the `admissible' 
Boolean sublattices of $\frak U$ the decoherence functional 
defines a partial logical implication which allows to make logical 
inferences. \item The description of a physical system and 
reasoning in terms of elements of $\cal U$ (homogeneous effect 
histories of the first kind) is only indirectly possible by using 
the map $\tau : {\cal U} \to {\frak U}$. 
\item While homogeneous effect histories have a direct physical 
interpretation in terms of time sequences of physical properties, 
inhomogeneous (effect) histories have no such direct 
interpretation. We tentatively suggest, however, that they may be 
interpreted as representatives of {\em unsharp quantum events}, 
i.e, events which cannot be associated with some fixed time, but 
which are smeared out in time. \end{itemize} \end{enumerate}

\subsubsection*{Acknowledgments}
I would like to thank Professor Frank Steiner for his 
support of my work. Financial support given by 
Deutsche Forschungsgemeinschaft (Graduiertenkolleg f\"ur 
theoretische Elementarteilchenphysik) is also gratefully 
acknowledged. 


\begin{thebibliography}{99}
\bibitem{Bell87} {\bf J.S.~Bell}, {\em  Speakable and 
Unspeakable in Quantum Mechanics} (Cambridge University Press, 
1987).
\bibitem{Bohm87} {\bf D.~Bohm, B.J.~Hiley and P.N.~Kaloyerou}, 
{ Physics Reports} {\bf 144}, 321 (1987). 
\bibitem{Giuntini91} {\bf R.~Giuntini}, {\em Quantum Logic 
and Hidden Variables} (Bibliographisches Institut \& 
F.A.~Brockhaus AG, Mannheim, 1991). 
\bibitem{Ghirardi86} {\bf G.C.~Ghirardi, A.~Rimini and T.~Weber}, 
{ Physical Review D} {\bf 34}, 470 (1986).
\bibitem{Pearle86} {\bf P.~Pearle}, { Physical Review 
D} {\bf 33}, 2240 (1986).
\bibitem{Pearle89} {\bf P.~Pearle}, { Physical Review 
A} {\bf 39}, 2277 (1989).
\bibitem{Gisin89} {\bf N.~Gisin}, { Helvetica Physica 
Acta} {\bf 62}, 363 (1989). 
\bibitem{Nakano94} {\bf A.~Nakano and P.~Pearle}, { 
Foundations of Physics} {\bf 24}, 363 (1994).
\bibitem{Griffiths84} {\bf R.B.~Griffiths}, { Journal 
of Statistical Physics} {\bf 36}, 219 (1984).
\bibitem{Griffiths93} {\bf R.B.~Griffiths}, { 
Foundations of Physics} {\bf 23}, 1601 (1993).
\bibitem{Griffiths94} {\bf R.B.~Griffiths}, {\em A 
Consistent History Approach to the Logic of Quantum Mechanics}, to 
appear in the Proceedings of the Symposium on the Foundations of 
Modern Physics 1994 Helsinki, Finland. 
\bibitem{Griffiths95} {\bf R.B.~Griffiths}, {\em Consistent 
Quantum Reasoning}, preprint, quant-ph/9505009 (1995). 
\bibitem{Omnes88a} {\bf R.~Omn\`es}, { Journal 
of Statistical Physics} {\bf 53}, 893 (1988). 
\bibitem{Omnes88b} {\bf R.~Omn\`es}, { Journal of 
Statistical Physics} {\bf 53}, 933 (1988). 
\bibitem{Omnes88c} {\bf R.~Omn\`es}, { Journal 
of Statistical Physics} {\bf 53}, 957 (1988). 
\bibitem{Omnes89} {\bf R.~Omn\`es}, { Journal of 
Statistical Physics} {\bf 57}, 357 (1989). 
\bibitem{Omnes90} {\bf R.~Omn\`es}, { Annals of Physics 
(N.Y.)} {\bf 201}, 354 (1990). 
\bibitem{Omnes92} {\bf R.~Omn\`es}, { Reviews of 
Modern Physics} {\bf 64}, 339 (1992).
\bibitem{Omnes94} {\bf R.~Omn\`es}, {\em The 
Interpretation of Quantum Mechanics} (Princeton University Press, 
1994).
\bibitem{Omnes95} {\bf R.~Omn\`es}, { Foundations of 
Physics} {\bf 25}, 605 (1995).
\bibitem{Isham94} {\bf C.J.~Isham}, { Journal of 
Mathematical Physics} {\bf 35}, 2157 (1994). 
\bibitem{IshamL94} {\bf C.J.~Isham and N.~Linden}, 
{ Journal of Mathematical Physics} {\bf 35}, 5452 (1994).
\bibitem{IshamL95} {\bf C.J.~Isham and N.~Linden}, 
{ Journal of Mathematical Physics} {\bf 36}, 5392 (1995).
\bibitem{IshamLS94} {\bf C.J.~Isham, N.~Linden and 
S.~Schreckenberg}, { Journal of Mathematical Physics} 
{\bf 35}, 6360 (1994).
\bibitem{GellMann90a} {\bf M.~Gell-Mann and J.B.~Hartle}, 
in: {\em 
Proceedings of the 25th International Conference on High Energy 
Physics, Singapore, August 2-8, 1990}, 1303, edited by K.K.~Phua 
and Y.~Yamaguchi (World Scientific, Singapore, 1990). 
\bibitem{GellMann 90b} {\bf M.~Gell-Mann and J.B.~Hartle}, 
in: {\em 
Proceedings of the Third International Symposium on the 
Foundations of Quantum Mechanics in the Light of New Technology}, 
321, edited by S.~Kobayashi, H.~Ezawa, Y.~Murayama 
and S.~Nomura (Physical Society of Japan, Tokyo, 1990). 
\bibitem{GellMann90c} {\bf M.~Gell-Mann and J.B.~Hartle}, 
in: {\em 
Complexity, Entropy and the Physics of Information, Santa Fe 
Institute Studies in the Science of Complexity}, Vol.~VIII, 425, 
edited by W.~Zurek (Addison-Wesley, Reading, 1990). 
\bibitem{GellMann93} {\bf M.~Gell-Mann and J.B.~Hartle}, 
{ Physical Review D} {\bf 47}, 3345 (1993).
\bibitem{GellMann94} {\bf M.~Gell-Mann and J.B.~Hartle}, 
{\em Equivalent Sets of Histories and Multiple 
Quasiclassical Domains}, preprint, 
gr-qc/9404013 (1994). 
\bibitem{GellMann95} {\bf M.~Gell-Mann and J.B.~Hartle}, 
{\em Strong Decoherence}, preprint, gr-qc/9509054 (1995). 
\bibitem{Hartle91} {\bf J.B.~Hartle}, in: {\em Quantum 
Cosmology and Baby Universes: Proceedings of the 1989 Jerusalem 
Winter School for Theoretical Physics}, 65, edited by S.~Coleman, 
J.B.~Hartle, T.~Piran and S.~Weinberg (World Scientific, 
Singapore, 1991).
\bibitem{Hartle94} {\bf J.B.~Hartle}, in: {\em Proceedings 
of the 1992 Les Houches Summer School,} B.~Julia and J.~Zinn-
Justin (eds.), Les Houches Summer School Proceedings Vol.~LVII 
(North Holland, Amsterdam, 1994).
\bibitem{Dowker96} {\bf F.~Dowker and A.~Kent}, { 
Journal of Statistical Physics} {\bf 82}, 1575 (1996). 
\bibitem{Kent95} {\bf A.~Kent}, {\em Remarks on Consistent 
Histories and Bohmian Mechanics}, to appear in: {\em Bohmian 
Mechanics and Quantum Theory: An Appraisal}, edited by J.~Cushing, 
A.~Fine and S.~Goldstein (Kluwer Academic Press); 
quant-ph/9511032 (1995).
\bibitem{Kent96} {\bf A.~Kent}, {\em Consistent Sets 
Contradict}, DAMTP/96-18, gr-qc/9604012 (1996). 
\bibitem{Zehn.d.} {\bf H.D.~Zeh}, {\em The Program of 
Decoherence: Ideas and Concepts}, intended as Chapter II in: 
D.~Guilini, E.~Joos, C.~Kiefer, J.~Kupsch, I.-O.~Stamatescu and 
H.D.~Zeh, {\em Decoherence and the Emergence of a Classical 
World} (Springer Verlag, Berlin, in preparation); also 
quant-ph/9506020.
\bibitem{Rudolph96} {\bf O.~Rudolph}, { International 
Journal of Theoretical Physics}, {\bf } forthcoming issue (1996), 
quant-ph/9512024. 
\bibitem{Busch89} {\bf P.~Busch, M.~Grabowski and P.J.~Lahti}, 
{ Foundations of Physics Letters} {\bf 2}, 331 (1989). 
\bibitem{Busch91} {\bf P.~Busch, P.J.~Lahti and P.~Mittelstaedt}, 
{\em The Quantum Theory of Measurement}, Lecture Notes in 
Physics {\bf m2} (Springer Verlag, Berlin, 1991).
\bibitem{Busch95} {\bf P.~Busch, M.~Grabowski and 
P.J.~Lahti}, {\em Operational Quantum Physics}, Lecture 
Notes in Physics {\bf m31} (Springer Verlag, Berlin, 1995).
\bibitem{Ludwig72} {\bf G.~Ludwig}, {\em Einf\"uhrung in 
die Grundlagen der Theoretischen Physik}, Volumes 1-4 (Vieweg, 
Braunschweig, 1972-79).
\bibitem{Kraus83} {\bf K.~Kraus}, {\em States, Effects, and 
Operations}, Lecture Notes in Physics vol.~{\bf 190} (Springer 
Verlag, Berlin, 1983). 
\bibitem{Wright95} {\bf J.D.M.~Wright}, { Journal of 
Mathematical Physics} {\bf 36}, 5409 (1995).
\bibitem{Bunce92} {\bf L.J.~Bunce and J.D.M.~Wright}, { 
Bulletin of the American Mathematical Society} {\bf 26}, 288 
(1992).
\bibitem{Bunce94} {\bf L.J.~Bunce and J.D.M.~Wright}, { 
Journal of the London Mathematical Society (2)} {\bf 49}, 133 
(1994).
\bibitem{Davies76} {\bf E.B.~Davies}, {\em Quantum 
Theory of Open Systems} (Academic Press, London, 1976).
\bibitem{Foulis94} {\bf D.J.~Foulis and M.K.~Bennett}, { 
Foundations of Physics} {\bf 24}, 1331 (1994). 
\bibitem{Kopka94} {\bf F.~K\^opka and F.~Chovanec}, { 
Mathematica Slovaca} {\bf 44}, 21 (1994).
\bibitem{Kopka92} {\bf F.~K\^opka}, { Tatra Mountains 
Mathematical Publications} {\bf 1}, 83 (1992). 
\bibitem{Dvurecenskij95} {\bf A.~Dvure\v{c}enskij}, { 
Transactions of the 
American Mathematical Society} {\bf 347}, 1043 (1995). 
\bibitem{DvurecenskijP94a} {\bf A.~Dvure\v{c}enskij and 
S.~Pulmannov\'a}, { 
International Journal of Theoretical Physics} {\bf 33}, 819 
(1994). 
\bibitem{DvurecenskijP94b} {\bf A.~Dvure\v{c}enskij and 
S.~Pulmannov\'a}, { 
Reports on Mathematical Physics} {\bf 34}, 151 (1994).
\bibitem{DvurecenskijP94c} {\bf A.~Dvure\v{c}enskij and 
S.~Pulmannov\'a}, { 
Reports on Mathematical Physics} {\bf 34}, 251 (1994).
\bibitem{Pulmannova95} {\bf S.~Pulmannov\'a}, { 
International Journal of Theoretical Physics} {\bf 34}, 189 
(1995). 
\bibitem{Langer62} {\bf H.~Langer}, { Acta Mathematica 
Academiae Scientiarum Hungaricae} {\bf 13}, 415 (1962). 
\end{thebibliography}
\end{document}